# Optical Probing of the Spin Polarization of the $\nu = 5/2$ Quantum Hall State

M. Stern[1],* P. Plochocka[2], V. Umansky[1], D. K. Maude[2], M. Potemski[2], and I. Bar-Joseph[1]
[1] *Department of Condensed Matter Physics, The Weizmann Institute of Science, Rehovot, Israel* and
[2] *Laboratoire National des Champs Magnétiques Intenses, CNRS-UJF-UPS-INSA, Grenoble, France*
(Dated: May 18, 2010)

We apply polarization resolved photoluminescence spectroscopy to measure the spin polarization of a two dimensional electron gas in perpendicular magnetic field. In the vicinity of filling factor $\nu = 5/2$, we observe a sharp discontinuity in the energy of the zero Landau level emission line. We find that the splitting between the $\sigma^+$ and $\sigma^-$ polarizations exhibits a sharp drop at $\nu = 5/2$ and is equal to the bare Zeeman energy, which resembles the behavior at even filling factors. We show that this behavior is consistent with filling factor $\nu = 5/2$ being unpolarized.

Since its discovery more than two decades ago [1], the fractional quantum Hall (FQH) state at filling factor $\nu = 5/2$ has been raising fundamental questions that challenges our understanding of strongly correlated two dimensional electron systems (2DES). One of the intriguing theories to describe this even denominator state was suggested by Moore and Read [2, 3]. The unique feature in this theory is that its elementary excitations exhibit non abelian statistics [4]. It was shown that such a non abelian state could be a good candidate for the realization of a topological quantum computer [5], which triggered considerable experimental and theoretical interest. Recent measurements of the $e/4$ quasiparticle charge at $\nu = 5/2$ [6], as well as the tunneling spectra [7], are consistent with the Moore-Read theory. However, an unambiguous and direct experimental evidence for non abelian statistics is still missing, and other less exotic abelian wavefunctions - such as the Halperin (3,3,1) state [8] - could also fit with the current experimental data.

A key feature of the $\nu = 5/2$ state that could help in constructing the appropriate wave function and test the relevance of the Moore-Read theory is the electron spin polarization. The Moore-Read theory explicitly assumes a spin polarized state and this property has been confirmed by an exact numerical diagonalization performed by Morf [9] which finds a fully polarized ground state. However, the current experimental evidence gathered from tilted field measurements [10, 11] seems to be inconsistent with this assumption and point to a spin unpolarized state. The importance of this issue for the understanding of the $\nu = 5/2$ state calls for further experimental investigations, based on different techniques and points of view.

In this work we apply polarization resolved photoluminescence (PL) spectroscopy to measure the spin polarization of the 2DES. The $\nu = 5/2$ state is clearly observed in the PL data as a sharp discontinuity in the energy of the zero Landau level (LL$_0$) emission line. We find that the energy splitting between the $\sigma^+$ and $\sigma^-$ emission lines exhibits a drop at $\nu = 5/2$ and is equal to the bare Zeeman splitting, which resembles the behavior at even filling factors. We show that this behavior is consistent with the $\nu = 5/2$ being a spin unpolarized state.

An essential ingredient in our measurement is the quality of the GaAs/AlGaAs heterostructure. The sample consists of a single 30-nm-wide GaAs/Al$_{0.25}$Ga$_{0.75}$As quantum well (QW) located 160 nm below the surface and doped on both sides with Si delta doping. The two delta doping were placed in narrow quantum wells, separated from the QW by an undoped Al$_{0.25}$Ga$_{0.75}$As layer of 80 nm thickness [12]. The sample was optimized for transport measurements in darkness, and it was clear that light illumination would change its properties. Hence, gating of the sample was essential in order to restore the electron density, and more importantly, the mobility. This was achieved by depositing a 4-nm PdAu semitransparent gate on the surface of the sample. The measurements were performed in a dilution fridge at a base temperature of 45 mK with a magnetic field applied along the growth axis of the wafer. The light source was a Ti:Sapphire laser at 720 nm and the sample was illuminated through a thick optical fiber at extremely low power densities $\leq 3\,\mu$ W/cm$^2$. The PL signal was collected by the same optical fiber through a circular polarizer. The wafer was processed into a Hall bar such that transport measurements could be performed simultaneously using a standard lock-in technique at 10.66 Hz and excitation current of 2 nA.

To characterize the sample we first performed conductivity measurement in darkness after illumination. In Fig. 1(a) we show the transverse and longitudinal resistivity, $\rho_{xy}$ and $\rho_{xx}$, as a function of magnetic field at gate voltage $V_g = -0.2$ V. The density of the 2DES $n_e$ at this gate is found by Hall measurement and Shubnikov de Haas (SdH) oscillations to be $n_e = 2.6 \times 10^{11}$ cm$^{-2}$ with a mobility of $\mu = 20 \times 10^6$ cm$^2$ V$^{-1}$ s$^{-1}$. We observe five significant fractions between $\nu = 2$ and 3; 11/5, 7/3, 5/2, 8/3 and 14/5. The $\nu = 5/2$ fraction is very well resolved, and one can clearly see a sharp minimum of the longitudinal resistivity $\rho_{xx}$ and a plateau at $\rho_{xy} = 0.4\, h/e^2$. The solid lines in Figs. 1(b) and (c) show the behavior of the mobility and density as a function of $V_g$. It is seen that the density can be tuned between 2 to $4 \times 10^{11}$ cm$^{-2}$ while the mobility of the system changes between 10 to $25 \times 10^6$ cm$^2$ V$^{-1}$ s$^{-1}$. The sharp decrease of the mobility around $V_g = 0$ V characterizes the onset

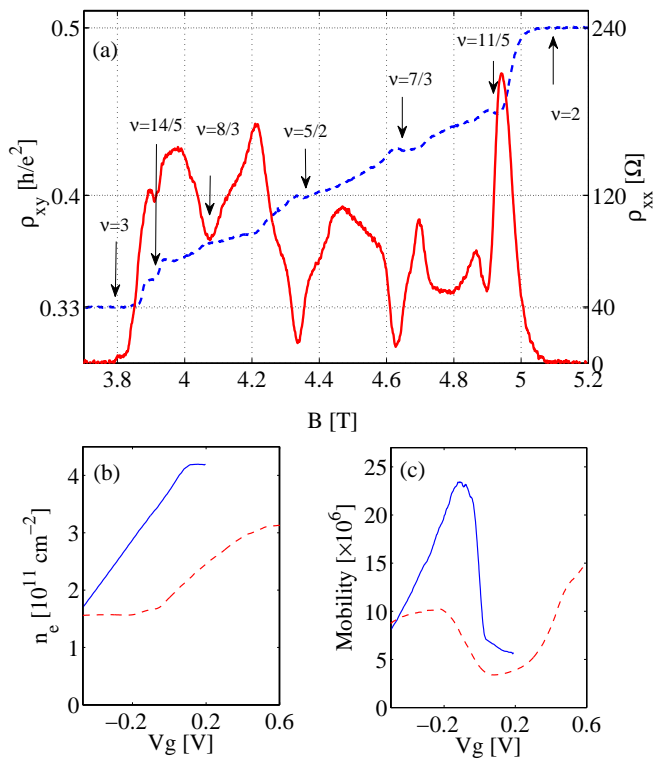

Figure 1: (a) Hall resistance $\rho_{xy}$ (blue dashed line) and longitudinal square resistance $\rho_{xx}$ (red solid line) as a function of the applied magnetic field $B$ measured in darkness at gate voltage $V_g = -0.2$ V. (b)-(c) Density and Mobility of the 2D electron gas as a function of $Vg$ in darkness (blue solid line) and with laser illumination at $\lambda = 720$ nm (red dashed line).

of the occupancy of the second QW subband.

We then performed the same measurements under laser illumination of the sample [dashed lines in Figs. 1(b) and (c)]. It is seen that the laser illumination depletes the QW, but this depletion can be easily compensated by applying a gate voltage. The gate voltage regime $0.4 < V_g < 0.6$ V seems to be optimized for the measurement, with $n_e \simeq 3 \times 10^{11}$ cm$^{-2}$ and $\mu \simeq 15 \times 10^6$ cm$^2$V$^{-1}$s$^{-1}$. A drawback of working with high positive gate voltage is the onset of a photoinduced leakage current between the gate and the 2DES. This leakage current interfered with the transport measurements and the quantum Hall data was of lower quality: the $\rho_{xx}$ dip at $\nu = 5/2$ was less pronounced and did not go all the way to zero, and the plateau was not well resolved. Nevertheless, the SdH oscillations and the Hall measurements allowed a precise determination of the electron density under illumination.

Figure 2(a) shows a compilation of the PL spectra at $V_g = 0.4$ V as $B$ is varied between $-6$ and $+6$ T. The intensity of the PL is color-coded, with dark red (blue) indicating strong (weak) signal. Under applied magnetic field, the emission spectrum of the 2DES forms a Landau level fan. Each of the Landau levels splits into a spin doublet, which can be resolved by analyzing the circular polarization of the PL. For $B < 0$, a valence hole with $J_z = -3/2$ and an electron with $S_z = +1/2$ recombine and emit a $\sigma^-$ photon. Similarly, a valence hole with $J_z = +3/2$ and an electron with $S_z = -1/2$ recombine and emit a $\sigma^+$ photon. This assignment of the light circular polarization is reversed at $B > 0$, and, hence, by fixing the circular polarizer and reversing the direction of the magnetic field we can measure each transition separately. In the following we shall concentrate on the main PL line, which is due to the recombination of a valence band hole with an electron from the lowest conduction band Landau level (LL$_0$). It is seen that this emission line exhibits sharp discontinuities for both polarizations. These discontinuities have been studied in recent decades both experimentally [13, 14] and theoretically [15–18], at integer and fractional filling factors.

In Figs. 2 (b) and (c) we show the PL spectra in the vicinity of $\nu = 5/2$. A clear and abrupt jump of the LL$_0$ PL energy is seen at $5/2$ ($B = 5.2$ T). This jump is more visible in the $\sigma^+$ polarization, where it is manifested as a change of the spectrum from a single line at lower field, to a doublet, and then again to a single line at higher field. This abrupt change is robust, is observed at different the gate voltages, and remains visible (but less pronounced) when the temperature is increased to 170 mK.

The photon energy in a PL experiment is given by the difference between the energy of the initial and final state of the system, $E_{PL} = E_f - E_i$. The initial state consists of the 2DES at its ground state and a valence band hole, while in the final state the valence hole disappears and the 2DES contains a quasi-hole. The understanding of this system is greatly simplified if one considers the *difference* between the emission energies at the two polarizations, $\Delta E = E_{PL}(\sigma^+) - E_{PL}(\sigma^-)$, rather than each of these energies separately [19, 20]. This energy difference, which we refer to as the *PL spin splitting*, factors out the contributions which are equal for the two spin polarizations, such as the cyclotron energy, the electron-hole direct Coulomb interaction in the initial state, and the electron-quasi-hole direct interaction in the final state.

It is useful to rewrite the *PL spin splitting* as $\Delta E = [E_f(\sigma^+) - E_f(\sigma^-)] - [E_i(\sigma^+) - E_i(\sigma^-)]$ and consider the initial and final state terms separately. Let us consider first the initial state terms. Since the density of valence band holes is extremely low and the electron-valence hole exchange interaction is very small, one can neglect the many body contributions and obtain $E_i(\sigma^-) - E_i(\sigma^+) = g_h \mu_B B$ where $g_h$ is the heavy-hole bare Landé factor. The final state is schematically described in Fig. 3. As can be seen the interaction energy in the final state strongly depends on the 2DES spin polarization. When the system is polarized the interaction energy of the quasi-hole with the electrons at the highest Landau level depends on the quasi-hole spin. Hence, we should get different interaction energies at $\sigma^-$ and $\sigma^+$. On the other hand, when the system is unpolarized the interaction of the quasi-hole with the 2DES is exactly the same for the two realizations, and the interaction energies at $\sigma^-$ and $\sigma^+$ are the same.

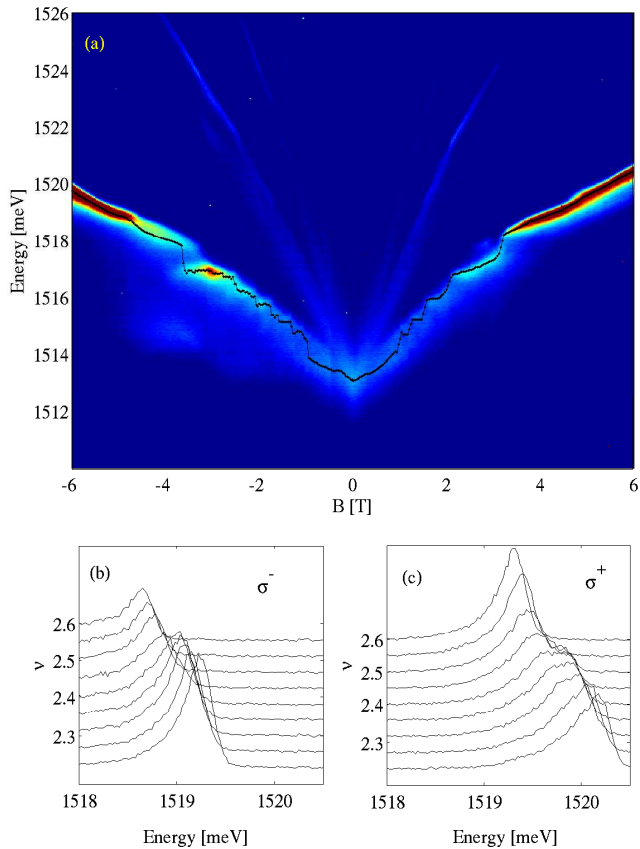

Figure 2: (a) PL spectra as a function of magnetic field $B$ for $V_g = 0.4$ V and $n_e = 2.85 \times 10^{11}$ cm$^{-2}$. The solid line fits the $LL_0$ PL energy. (b-c) Spectra in the vicinity of $\nu = 5/2$ for both polarization $\sigma^+$ and $\sigma^-$ for $V_g = 0.6$ V and $n_e = 3.15 \times 10^{11}$ cm$^{-2}$.

If we separate the bare Zeeman energy $g_e\mu_B B$ from the interaction term $\Delta\Sigma$ we can write the PL spin splitting as:

$$\Delta E = (g_h + g_e)\mu_B B + \Delta\Sigma \quad (1)$$

where $\Delta\Sigma$ describes the difference between the interactions of the quasi-hole in the lowest Landau level with the electron sea in $\sigma^-$ and $\sigma^+$. This term was actually calculated several decades ago by Ando [21], in the context of the enhanced g factor of a 2DES in a magnetic field. It was shown there that the term is oscillatory in $\nu$, and can be written as $\Delta\Sigma = E_0 \times (n_\uparrow - n_\downarrow)$ where $n_\uparrow - n_\downarrow$ is the spin polarization. An approximate value for $E_0$ can be obtained using the Hartree Fock approximation and neglecting the screening of the Coulomb interaction. Under these assumptions $E_0$ is the exchange energy $e^2/\varepsilon l_B$, where $l_B$ is the magnetic length and $\varepsilon$ is the dielectric constant. One can see that $\Delta\Sigma = 0$ when $n_\uparrow - n_\downarrow = 0$, in accordance with Fig. 3. Clearly, a more realistic model is needed to correctly account for screening and electron-hole correlations in the final state [17, 22] if one wishes to quantitatively relate the spin polarization to the values of the observed splitting. Nevertheless, the fact that $\Delta\Sigma$ vanishes when the initial state is spin depolarized should remain.

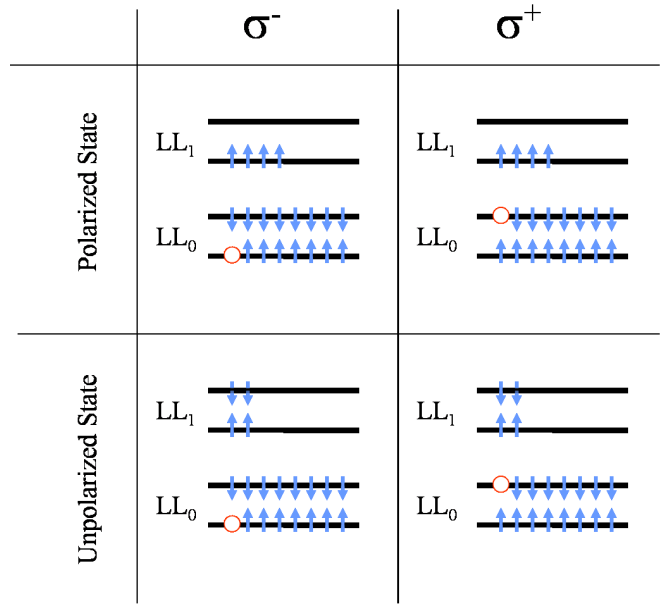

Figure 3: A schematic representation of the final state at both polarizations for a $\nu = 5/2$ spin polarized and unpolarized state. The up and down arrows correspond to electrons with spin up and down respectively in the two first Landau levels. The dots correspond to the remaining quasihole in $LL_0$.

Let us turn now to the experimental results for the *PL spin splitting*. We first determined the energy difference, $\Delta E$, between the peak positions of the PL lines at the two polarizations at each magnetic field. The precision of this procedure was very good at high field (see Figs. 4 a,b) and deteriorated at low fields, where the emission line was broad; the signal to noise ratio for $\Delta E$ is 25 at $B = 5$ T and reduces to 1 at $B = 1.5$ T. In cases where the spectrum consists of a doublet (around $\nu = 5/2$ and 3) the emission energy was calculated as the center of mass of the two lines. We find that the *PL spin splitting* curve $\Delta E(B)$ has an oscillatory component which is superimposed on a constant slope of 0.12 meV/T, which is the bare Zeeman splitting, $(g_h + g_e)\mu_B B$ (Eq. 1). Figure 4 (c) shows the resulting $\Delta\Sigma$ as a function of filling factor for $V_g = 0.4$ and 0.6 V. It is seen that $\Delta\Sigma$ is semi-periodic in $\nu$, with a period of $\Delta\nu = 2$, and minima at $\nu = 2, 4, 6$ and 8 [19], in agreement with the Ando formula. Remarkably, we observe a clear dip of $\Delta\Sigma$ at $\nu = 5/2$; at this filling factor $\Delta\Sigma$ vanishes and the *PL spin splitting* is equal to the bare Zeeman splitting.

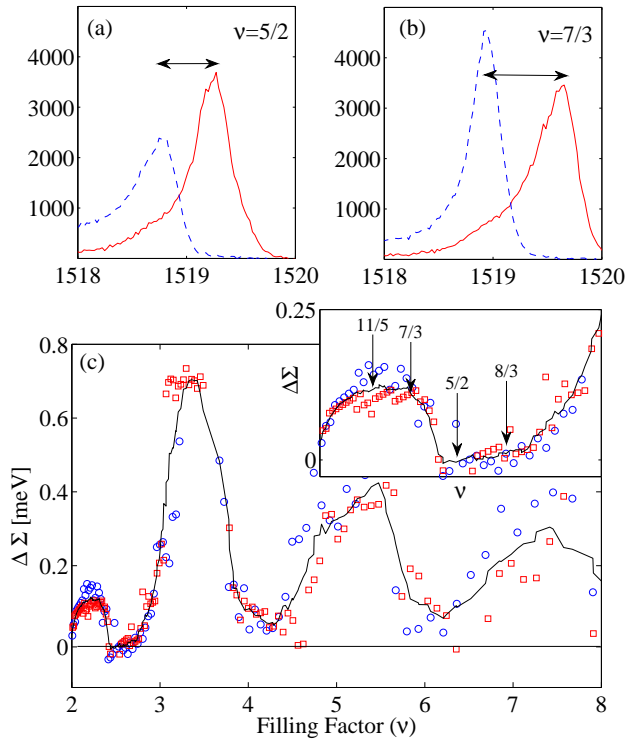

Figure 4: (a-b) Spectra at both polarizations for $\nu = 5/2$ and $\nu = 7/3$ respectively. The arrows show the difference in the PL spin splitting. (c)$\Delta\Sigma$ as a function of filling factor $\nu$ at $V_g = 0.4$ V (red square) and $V_g = 0.6$ V (blue circle). The black line is a guide for the eyes. (inset) Zoom in the region between $\nu = 2$ and $\nu = 3$.

We interpret this finding as an indication that the $\nu = 5/2$ state is unpolarized, $n_\uparrow - n_\downarrow = 0$. It is instructive to compare $\Delta\Sigma$ at 5/2 to that of the adjacent fraction, 7/3, the first exhibiting a minimum while the other - a maximum. This difference suggests that the spin polarization of the two fractions is different, 5/2 unpolarized and 7/3 polarized. This analysis also indicates that 8/3 is unpolarized (inset of Fig. 4 c).

A valid question is related to the effect of the illumination on the 2DES, primarily the creation of a steady state density of valence band holes and of quasi-holes in the electron gas. At the low illumination levels used in our experiment ($\leq 3\mu$ W/cm$^2$) the estimated steady state valence band hole density is extremely small; assuming a recombination time of 1 ns we should obtain a density of $\sim 3 \times 10^2$ cm$^{-2}$, which can safely be neglected. To estimate the steady state density of the quasi-holes one needs the relaxation time of the 2DES to the ground state after recombination. Taking this relaxation time to be $\sim 10^{-7}$ seconds [23], one gets a steady state quasi-holes density of $\sim 3 \times 10^5$ cm$^{-2}$, which is 6 orders of magnitude lower than the electron density. This small density corresponds to a net increase of the 2DES temperature by 0.2 mK, and one can therefore safely neglect this effect as well.

In conclusion, the PL data suggests that the 5/2 state is spin-unpolarized. This observation puts a tight constraint on the type of wavefunction that could describe this state. Our data is inconsistent with the Moore-Read theory [2], which assumes a spin polarized state, and with the results of numerical calculations [9]. An intriguing question is whether one can construct a non-abelian theory assuming a spin-unpolarized ground state.

We thank Ady Stern, N. Read, M. Dolev, R. Ilan and B. Piot for fruitful discussions. This research was supported by the Israeli Science Foundation and by the Minerva Foundation. P. P is financially supported by the EU under FP7, contract no. 221249 'SESAM'.